\documentclass{appolb}
\usepackage{graphicx}
\usepackage{hyperref}
\usepackage{amsmath}
\usepackage{amssymb}
\usepackage[numbers,sort&compress]{natbib} 

\begin{document}
\title{On Dyson-Schwinger studies of Yang-Mills theory and the four-gluon vertex
\thanks{Presented at Excited QCD 2015, 8-14 March 2015, Tatranska Lomnica, Slovakia}
}

\author{Markus Q. Huber$^1$, Anton K. Cyrol$^2$, Lorenz von Smekal$^{3}$
\address{$^1$Institute of Physics, University of Graz, NAWI Graz, Universit\"atsplatz 5, 8010 Graz, Austria}
\\
\address{$^2$Institut f\"ur Theoretische Physik, Ruprecht-Karls-Universit\"at Heidelberg, 69120 Heidelberg, Germany}
\address{$^3$Institut f\"ur Kernphysik, Theoriezentrum, Technische Universit\"at Darmstadt, 64289 Darmstadt, Germany}
}

\maketitle
\begin{abstract}
We review the status of calculations of Yang-Mills Green functions from Dyson-Schwinger equations. The role of truncations is discussed and results for the four-gluon vertex are presented.
\end{abstract}

\section{Truncations of functional equations}

Functional equations provide a non-perturbative method applicable to the study of the strong coupling regime of quantum chromodynamics (QCD). 
They are organized in systems of equations that each provide a complete description of QCD. However, since these systems are infinitely large, they have to be truncated to a manageable subset. Thereby, the notion of 'manageable' has changed over time with an increasing number of correlation functions considered within a given truncation. Still, typically some model input is required . The goal would be, of course, to solve a system that does not rely on models but still provides a good quantitative description.

One challenge of such studies is the estimation of the size of the truncation effects. Unfortunately, there is no easy and generic way to quantify them. In the following we discuss this issue for the case of Dyson-Schwinger equations (DSEs).

The effect of truncating a DSE depends a lot on the energy regime considered. For large momenta it is obvious that perturbation theory provides an estimate of how good a truncation is. In the non-perturbative regime, on the other hand, ordering schemes are more elusive but sometimes known. For example, in the deep IR, the full towers of DSEs and functional flow equations can be analyzed \emph{without} truncations and the leading diagrams can be identified if the scaling type solutions \cite{vonSmekal:1997is} of Yang-Mills theory in the Landau \cite{Fischer:2009tn} or the maximally Abelian gauges \cite{Huber:2009wh} are considered.  However, even in this case, the effect of a truncation in the mid-momentum regime is difficult to estimate, as we have no scheme that tells us about the importance of single ingredients.

The easiest possibility for checking the reliability of a truncation is the comparison with other methods. This is extensively done with results from lattice calculations -- where they are available -- and allowed, for instance, identifying the importance of two-loop contributions in the gluon propagator and the three-gluon vertex \cite{Blum:2014gna,Eichmann:2014xya}: They are important for the propagator, but not so much for the vertex.
However, the comparison to lattice results is not always possible. For example, calculating correlation functions beyond three-point functions is currently not feasible on the lattice, and at non-zero densities the sign problem constitutes a serious obstacle.
Thus one has to turn to other possibilities. A straightforward one is the comparison with another truncation that contains additional correlation functions. However, it is not clear if adding more correlation functions beyond that might lead to large changes as well. Also, it is difficult to isolate the effect of a single change in the truncation, because although there might be quite some differences between the results from two truncations differing only by the inclusion of one correlation function, the importance of these additional contributions can only be estimated once 'parallel' effects are understood. For example, using a realistic ghost-gluon vertex does have an effect on the propagators \cite{Huber:2012kd,Aguilar:2013xqa}, but the employed model for the three-gluon vertex can still counteract them in the gluon propagator equation \cite{Huber:2012kd}.

Having no systematic expansion scheme to truncate DSEs, is it possible to exclude the possibility that higher correlation functions are even more important than the lower ones? We argue that there is at least a formal limit to the importance of higher correlation functions. It is based on the success of lattice simulations which are always done for a finite number of lattice sites $N$. Thus, a correlation function with more than $N$ legs cannot live on such a lattice and is consequently not contained in the simulation. Still it provides a reasonably good description. Of course, the typical number of lattice sites is vastly larger than any number of legs of a correlation function ever calculated, but the mere existence of such a number shows that not arbitrarily high correlation functions are required to obtain good results. Thus we conclude that at some point higher correlation functions can not be important anymore. The remaining question: How many correlation functions do we need to obtain a quantitative description?

The results of the last few years give us already a hint and encourage the view that the calculation of correlation functions within a truncation that leads to quantitatively good results is feasible. While previously new truncations sometimes led to a change in the qualitative behavior, the changes seen nowadays are all of a quantitative nature. For example, the effects of three-point functions on the propagators are by now better understood \cite{Huber:2012kd,Aguilar:2013xqa,Blum:2014gna} which was possible only by explicit calculations of the three-point functions \cite{Huber:2012kd,Pelaez:2013cpa,Aguilar:2013xqa,Blum:2014gna,Eichmann:2014xya}. An important finding was the relevance of the two-loop diagrams in the gluon propagator DSE \cite{Blum:2014gna}. However, it seems that in higher correlation functions two-loop diagrams are not as important. At least the calculation of the three-gluon vertex and its comparison with lattice results hints at this \cite{Blum:2014gna}.

Finally, also the four-gluon vertex was calculated \cite{Binosi:2014kka,Cyrol:2014kca,Cyrol:2014mt}. It appears, for instance, in the truncated DSE of the three-gluon vertex, where a model was used \cite{Blum:2014gna,Eichmann:2014xya}. The study of the effect of the four-gluon vertex on lower correlation functions will be an interesting work for the future. In \cite{Cyrol:2014mt} the coupling of the three- and four-gluon vertices was already investigated and only small changes were found. The important point of \cite{Binosi:2014kka,Cyrol:2014kca,Cyrol:2014mt} is, however, that there are no unexpected features present and the behavior deviates rather little from the tree-level behavior. We will discuss the calculational setup and the results of \cite{Cyrol:2014kca} for the four-gluon vertex  below.

\section{Four-gluon vertex}

\subsection{Truncation}

\begin{figure}
\centerline{
\includegraphics[width=0.87\textwidth]{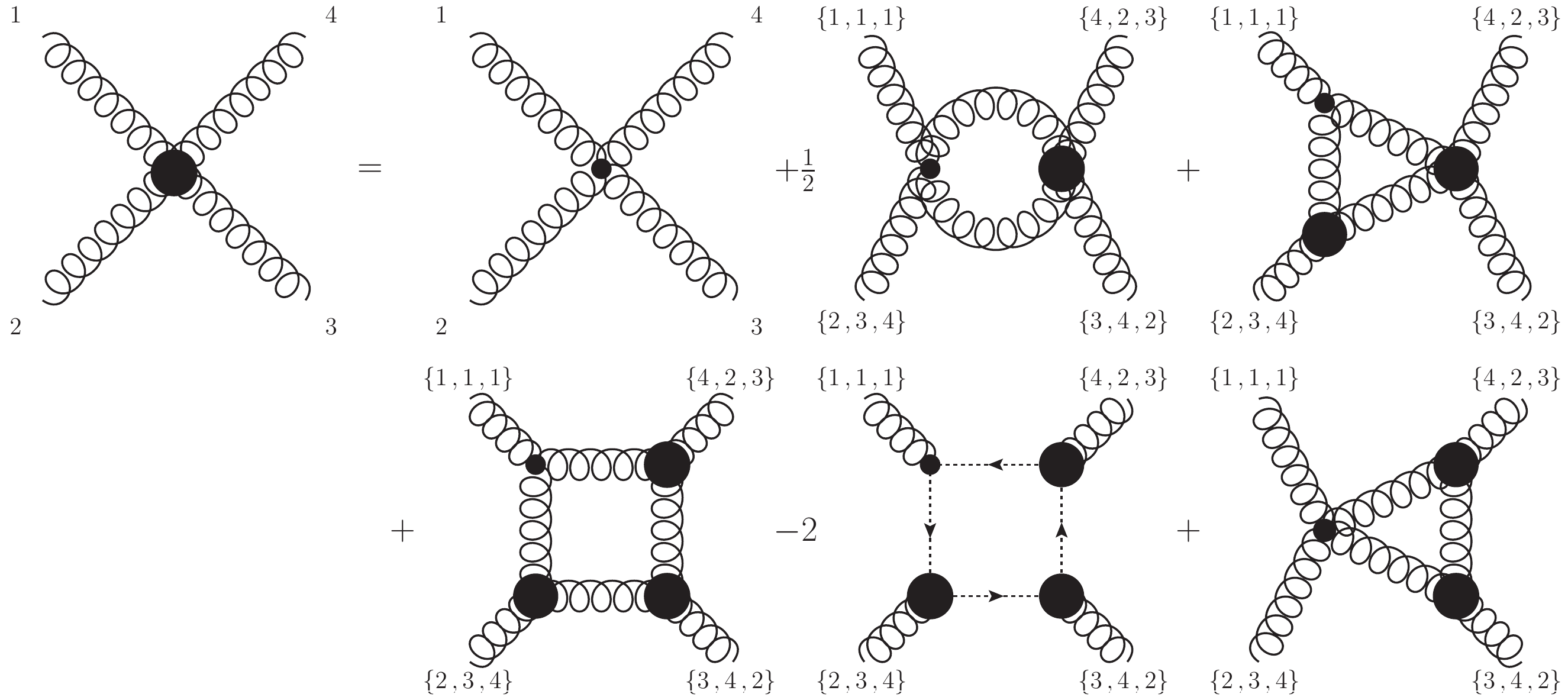}}
\caption{\label{fig:4g_DSE}Diagrammatic representation of the truncated four-gluon vertex. Dashed lines are ghosts, wiggly lines gluons. All internal propagators are dressed. The numbers at the legs refer to further diagrams obtained from permutations.}
\end{figure}

Three-point functions can be solved for all transverse dressing functions with full momentum resolution \cite{Huber:2012kd,Eichmann:2014xya}, but this is no longer possible for a four-point function. The reason is that the number of dressing functions as well as the numerical effort increase considerably. Motivated by the three-gluon vertex, where it is known that the reduction to the tree-level tensor leads to minor deviations only \cite{Eichmann:2014xya}, we reduce the vertex tensor structure similarly, $\Gamma^{abcd}_{\mu\nu\rho\sigma}(p,q,r,s)=\Gamma^{{(0)},abcd}_{\mu\nu\rho\sigma}D^\text{4g}(p,q,r,s)$, but keep the full momentum dependence.

The DSE of the four-gluon vertex, which we derived using \textit{DoFun} \cite{Huber:2011qr}, contains 60 diagrams. We retain all UV leading diagrams what amounts to discarding all diagrams that are either two-loop (39 diagrams) or feature a non-primitively divergent correlation function (5 diagrams). The truncated DSE is shown in Fig.~\ref{fig:4g_DSE}. The Bose symmetry of the vertex, which is broken by the truncation, is restored by explicit symmetrization. We note that since we have the full momentum dependence, only one diagram of each type needs to be calculated, because the symmetrized and crossed diagrams can be obtained from that.

The renormalization of the vertex is performed within the \textit{MiniMOM} scheme \cite{vonSmekal:1997is,vonSmekal:2009ae}. There, the renormalization constants of the propagators are fixed via a \textit{MOM} scheme and that of the ghost-gluon vertex via the \textit{MS} scheme. The renormalization constants for the three- and four-gluon vertices are determined by corresponding Slavnov-Taylor identities.

We consider here only the four-gluon vertex and treat all other correlation functions as given input. For the propagators we use the results of~\cite{Huber:2012kd}, which are in good agreement with available lattice results. This was possible by the use of an optimized effective three-gluon vertex that mimics the effects of two-loop diagrams. The same propagators were used for the calculation of the three-gluon vertex employed here \cite{Blum:2014gna}.
The ghost-gluon vertex, on the other hand, is taken as bare, since it does not have a large impact on the four-gluon vertex, as we showed explicitly in \cite{Cyrol:2014kca}.

\subsection{Results}

\begin{figure}
\centerline{
\includegraphics[width=0.45\textwidth]{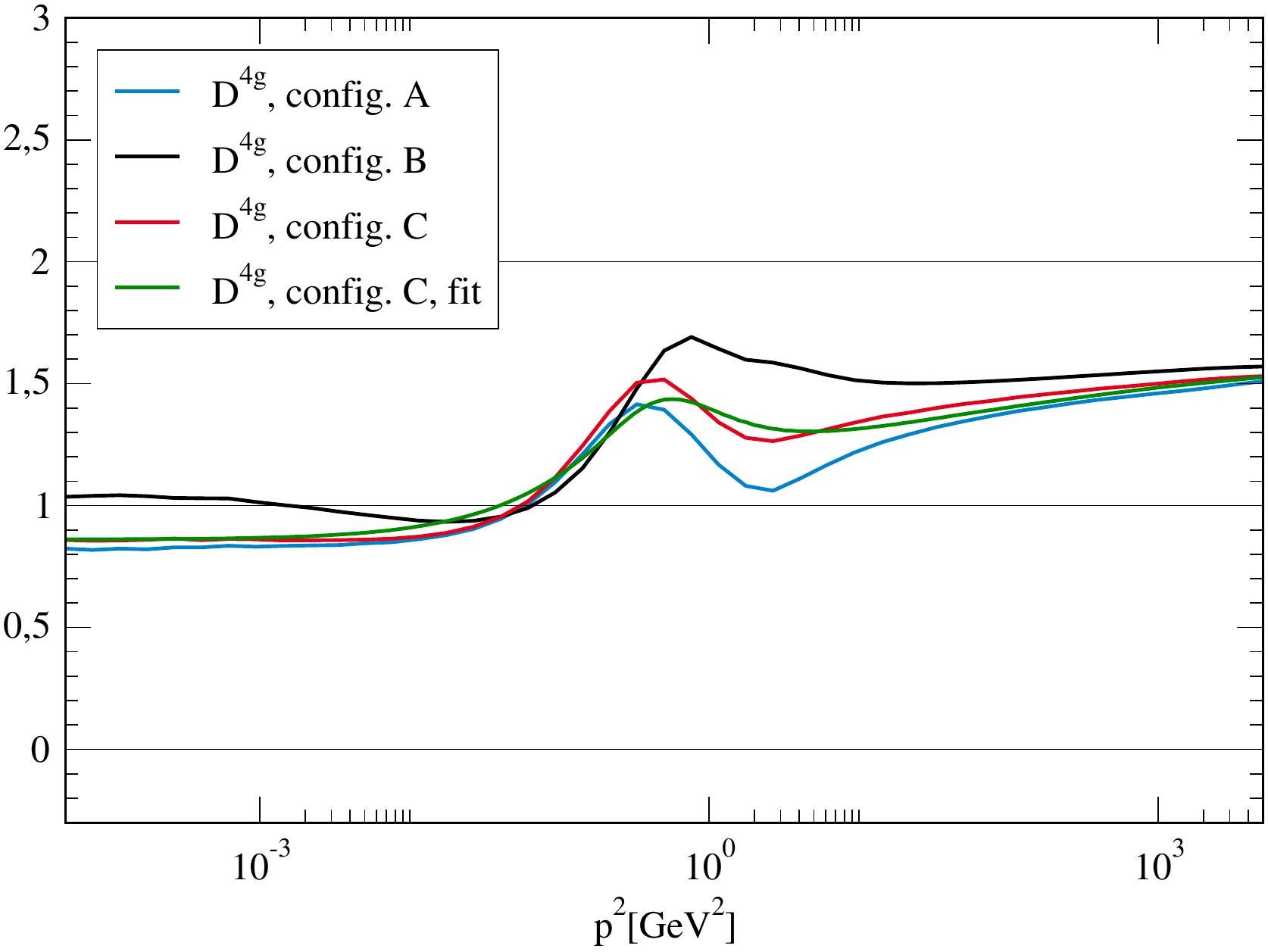}\hfill
\includegraphics[width=0.45\textwidth]{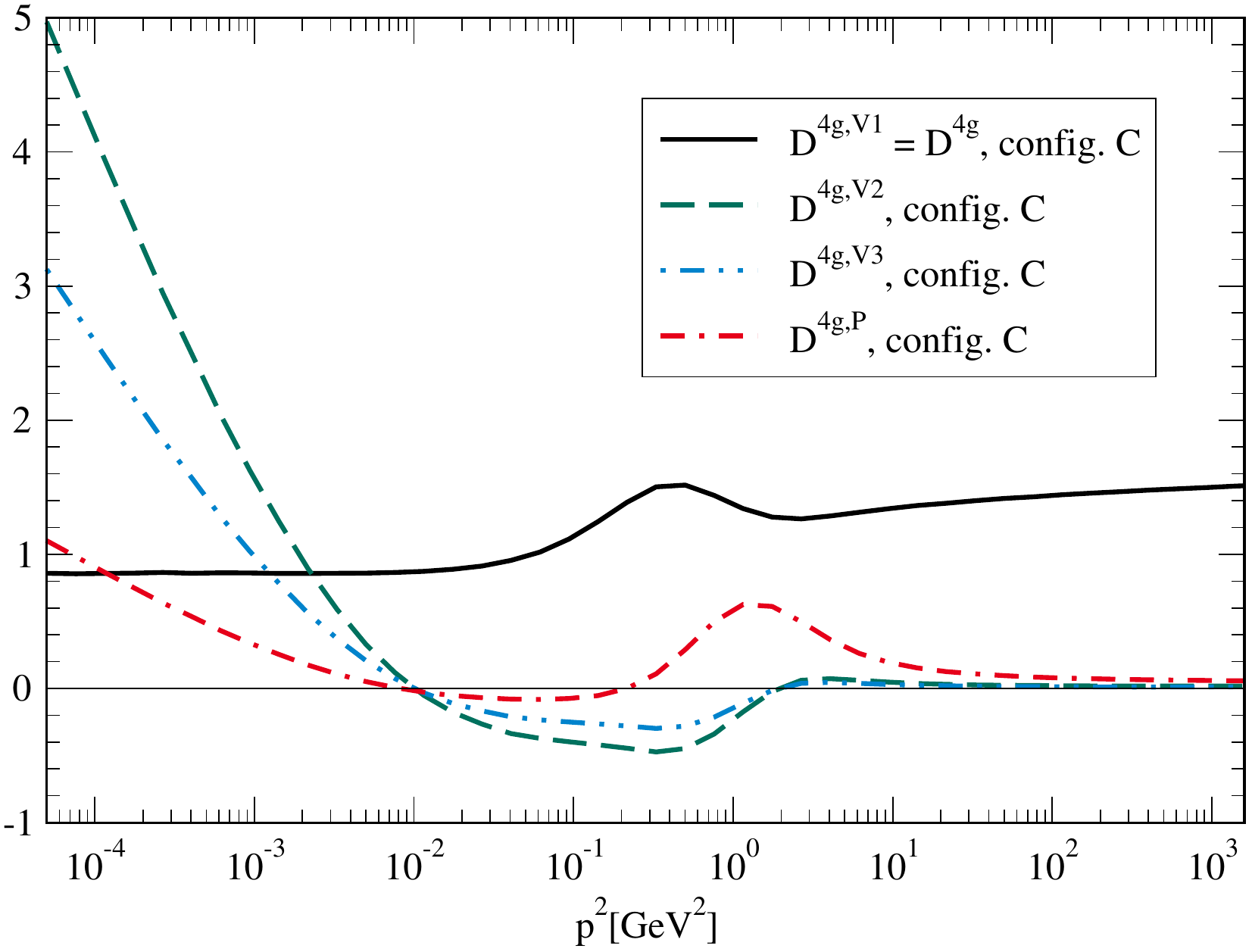}}
\caption{\label{fig:4g_results}Results for the four-gluon vertex. \textit{Left}: Certain momentum configurations and a fit.
\textit{Right}: Momentum dependence of different dressing functions.
}
\end{figure}

We solved the truncated four-gluon vertex DSE self-consistently employing the framework of \textit{CrasyDSE} \cite{Huber:2011xc} using a standard iterative procedure. For illustration purposes we chose three kinematic configurations $\boldsymbol{A}$, $\boldsymbol{B}$ and $\boldsymbol{C}$, see \cite{Cyrol:2014kca,Cyrol:2014mt} for details. The results for these configurations are plotted in Fig.~\ref{fig:4g_results} (left). They are qualitatively very similar. This is also true for other configurations, because the angle dependence is rather weak similar as for the three-gluon vertex \cite{Blum:2014gna}.

To assess the effect of truncating the vertex to its tree-level Lorentz and color structure, we calculated three more dressing functions using our previously obtained results as input. They were not obtained self-consistently, but judging from the already rather small corrections to the tree-level dressing function, we expect that the results for the other dressing functions respect the correct qualitative behavior. In contrast to the IR finite tree-level dressing function, we find a logarithmic IR divergence of these tensors as observed previously in~\cite{Binosi:2014kka}. However, the divergence sets in at very low momenta. As can be seen in Fig.~\ref{fig:4g_DSE} (right), the tree-level tensor is dominant except for very low momenta.\footnote{We checked that the difference to the results of \cite{Binosi:2014kka} in the mid-momentum regime stems from taking into account here the renormalization factor for each diagram.} It is important to note that the smallness of the non-tree-level dressing functions comes from cancellations between different diagrams which each have large contributions.

\section{Conclusions}

The four-gluon vertex was the last primitively divergent correlation function of Yang-Mills theory to be calculated self-consistently within a modern truncation scheme. The results are encouraging in the sense that no large corrections are found, except for IR divergences in non-tree-level dressing functions. However, they start at very low momenta and are only logarithmic, so that they are not expected to have a large impact. Of interest is the fact that within the employed truncation there is, in contrast to lower Green functions, no model dependence on higher Green functions. Even better, for the calculation of the two-loop diagrams in the gluon propagator, which are expected to yield important quantitative contributions, no correlation function beyond the four-gluon vertex is required. Thus the employed truncation provides a self-contained system that does not require any model input.

\bigskip

Funding by the FWF (Austrian science fund) under Contract P 27380-N27 is gratefully acknowledged.

\bibliographystyle{utphys_mod}
\bibliography{literature_eQCD2015_dses}

\end{document}